\documentclass[]{spie}  

\usepackage{amsmath,amsfonts,amssymb}
\usepackage{graphicx}
\usepackage[colorlinks=true, allcolors=blue]{hyperref}
\usepackage{glossaries}

\newacronym{pcs}{ELT-PCS}{Planetary Camera and Spectrograph}
\newacronym{elt}{ELT}{Extremely Large Telescope}
\newacronym{hci}{HCI}{high-contrast imaging}
\newacronym{ifu}{IFU}{integral field unit}
\newacronym{ifs}{IFS}{integral field spectrograph}
\newacronym{ghost}{GHOST}{GPU-based high-order adaptive optics test bench}
\newacronym{slm}{SLM}{spatial light modulator}
\newacronym{ao}{AO}{adaptive optics}

\title{Comparing the contrast performance of two IFU technologies for exoplanet direct imaging with ELT-PCS}

\author[a]{Ryan Griffiths}
\author[a]{Siddharth Maharana}
\author[a]{Matthias Tecza}
\affil[a]{University of Oxford, Denys Wilkinson Building, Oxford, UK.}

\authorinfo{Further author information: (Send correspondence to R.G.)\\R.G: E-mail: ryan.griffiths@physics.ox.ac.uk}

\begin{document} 
\maketitle

\begin{abstract}
A key science goal for the \gls{pcs} instrument is to image rocky exoplanets in reflected light around nearby M- and K-dwarfs. These observations will require a contrast of $10^{-9}$ at 100 mas angular separation, and $10^{-8}$ at 10 mas. To achieve such an extreme performance, every instrument design choice must be carefully evaluated. For \gls{pcs}'s science instruments, perhaps the most critical design choice is which \gls{ifu} technology to use. To this end, we have developed a modular, lab-based integral field spectrograph test bench with which to compare the performance of two popular \gls{ifu} technologies: image slicers and lenslet arrays. The test bench implements a Lyot coronagraph and a spatial light modulator to simulate a high-contrast imaging system, which feeds each \gls{ifu} with a diffraction-limited PSF image. Each \gls{ifu} is coupled to the spectrograph, which achieves R$\approx$3000 with the image slicer and R$\approx$100 with the lenslet array. The calibration and data reduction for each \gls{ifu} design are described and the initial results of a contrast comparison are presented. The lenslet array achieves a deeper contrast at angular separations of less than $6 \lambda/D$. The image slicer, however, gains more at small separations from applying spectral deconvolution, with a 3.4$\times$ greater gain at $4\lambda/D$. Spectral deconvolution further yields a contrast curve close to the detector noise floor. Further work will focus on carrying out a fully-controlled characterisation of the lenslet array \gls{ifu} and a more comprehensive study into the effects of adaptive optics residuals on the contrast.

\end{abstract}


\glsresetall

\section{INTRODUCTION}
\label{sec:intro}  

The \gls{pcs} is a dedicated direct imaging instrument planned for the 39 m \gls{elt}. It will implement an extreme adaptive optics system, coronagraph, and array of imaging and spectroscopic instrumentation to detect and characterise rocky exoplanets in reflected light around nearby K- and M-dwarfs for the first time \cite{Kasper2021PCSELT}. To probe a sufficient number of targets, \gls{pcs} will aim to observe exoplanetary systems in the visible and near-infrared at contrasts of $10^{-9}$ at 100 mas angular separation, and $10^{-8}$ at 10 mas. Existing \gls{hci} instruments such as VLT-SPHERE achieve contrasts of up to $10^{-5}$ at their inner working angle and up to $10^{-7}$ at larger separations beyond 200 mas \cite{Beuzit2019SPHERE:Telescope}. Therefore, each subsystem of \gls{pcs} will need to push the limits of current technology to achieve its contrast goals.

At the time of writing, the leading instrument concepts for \gls{pcs} include at least two \glspl{ifu}: a near-infrared JHK band \gls{ifu} operating at mid-spectral resolving powers of $R\approx10^{4}$, and a high-resolution VRI \gls{ifu} with $R>10^{5}$. A low-resolution \gls{ifu} with $R~\approx 10^2$ is also under consideration for spectro-polarimetry. Given the extreme contrast requirements of \gls{pcs}, the choice of \gls{ifu} technology will be critical. The current suite of high-contrast \glspl{ifs} such as SPHERE \cite{Beuzit2019SPHERE:Telescope}, GPI \cite{Macintosh2014FirstImager} and CHARIS \cite{Currie2020On-skySystem}, are dominated by microlens-array \glspl{ifu}, which, in general, provide lower spectral resolving power due to reformatting constraints, but are simple to produce and can image a wide field. Upcoming \gls{elt} instruments with direct imaging capabilities such as METIS and HARMONI implement image slicers \cite{Absil2024METISTesting, Houlle2021DirectModule}. Slicers make more efficient use of the detector space, enabling a higher spectral resolving power, but are more complex to manufacture and align. Additional alternatives include single mode optical fibres, for example the MCIFU \cite{Haffert2020Multi-coreFirst-light} instrument planned for MagAO-X and the ANDES fibre \gls{ifu} \cite{Marconi2024ANDESDevelopments}, or even a combination of lenslets and image slicers as in SCALES \cite{Stelter2023TheSpectrograph}. 

Selecting the optimal \gls{ifu} technology will require experimental testing of real optical systems. To that end, we have designed a modular experiment\cite{Meyer2024AnIFUs} to compare the contrast achieved by the two most widely-implemented \glspl{ifu} technologies: lenslet arrays and image slicers. The experiment has been recently upgraded to include a high-contrast imaging simulation in its front-end, implementing a \gls{slm} and a Lyot coronagraph, in addition to feeder optics which deliver a slow, diffraction-limited point-source image onto the \gls{ifu}. The spectrograph and camera are made from spare components built for the SWIFT integral field spectrograph \cite{Tecza2006SWIFT:Spectrograph}, and the detector is a cooled, 2K E2V CCD. The experiment is modular with respect to the mounting of the \gls{ifu} such that the alignment of other optical elements is mostly preserved during comparison studies. Furthermore, the updated experiment design has been made with compatibility with the \gls{ghost} experiment at ESO in mind \cite{Engler2022GPU-basedTestbench}. After studies have concluded in Oxford, the experiment will travel to \gls{ghost} to be tested behind its two-stage cascade adaptive optics system and coronagraph. In this paper we will describe the optical components of the \gls{ifs}, the implementation of its high-contrast imaging simulator system, other major upgrades to the test bench  and preliminary findings from comparing the contrast of the two \glspl{ifu}. The effects of post-processing with spectral deconvolution are also explored on each system. 

\section{IFU Designs}

\subsection{Image Slicer}

\begin{figure}[h]
    \centering
    \includegraphics[width=\linewidth, trim = 0cm 0cm 0cm 0cm]{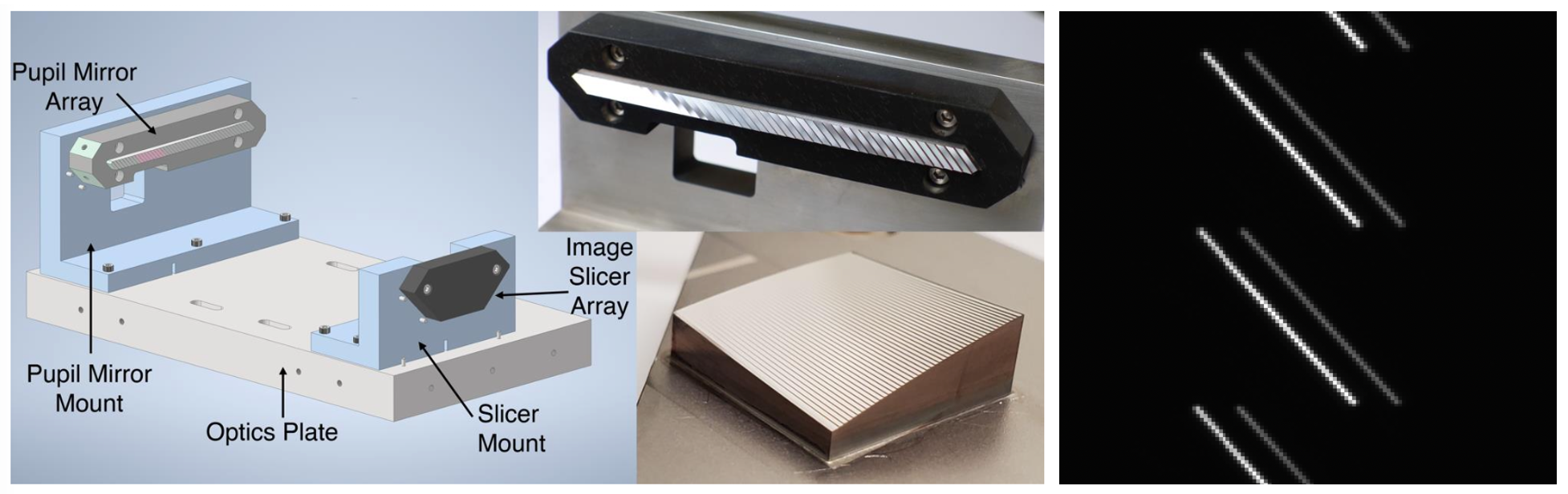}
    \caption{Image slicer \gls{ifu} design. The left image shows a CAD model of the two slicer mirror arrays mounted on an Invar plate. The centre images show the two mirror stacks as built by Canon. The image on the right is a sub-region of an arc-lamp flat field illumination of the slicer showing the slices imaged at 45$^\circ$ to the detector pixels. Images are partly reproduced from Ref.~\cite{Meyer2024AnIFUs}}
    \label{fig:slicerifu}
\end{figure}

The twisted image slicer was designed at the University of Oxford and manufactured by Canon Inc. in 2020 \cite{Tecza2022ImageELT-PCS}. In this design, a stack of parabolic slicing mirrors sits at 45$^\circ$ to the dispersion axis and detector pixels, which allows for spatial and spectral Nyquist sampling at the detector without the need for anamorphic optics typically required for square spaxels on-sky \cite{Tecza2014ImageOptics}. An image of the mounted slicer stack and pupil-mirror array is given in figure~\ref{fig:slicerifu} in addition to a CAD model of the overall assembly which includes a large Invar alignment plate. The image slicer optical parameters are summarised in table~\ref{tab:slicer}. We note that the pseudo-slit formed by the slicer is slightly longer than anticipated, and just over 44 slices fit on the CCD detector instead of the nominally expected 45 slices.

\begin{table}[ht]
\caption{Image slicer optical properties.}
\label{tab:slicer}
\centering
\begin{tabular}{ll}
\hline
\textbf{Parameter} & \textbf{Value} \\
\hline
Spaxel format & $45 \times 45$ spaxels\\
Spaxel size & 0.3 mm $\times$ 0.3 mm\\ 
Mirrors & parabolic (slicing) + spherical (pupil) \\
Exit pupil position & $\infty$, telecentric \\
Nominal Input & $f/850$, telecentric \\
Exit slit sampling & 2 pix \\
R & $\approx$ 3000 \\
Spectral Range & 720 - 916 nm\\
\hline
\end{tabular}
\end{table}

\subsection{Lenslet Array}

\begin{figure}[h]
    \centering
    \includegraphics[width=0.9\linewidth]{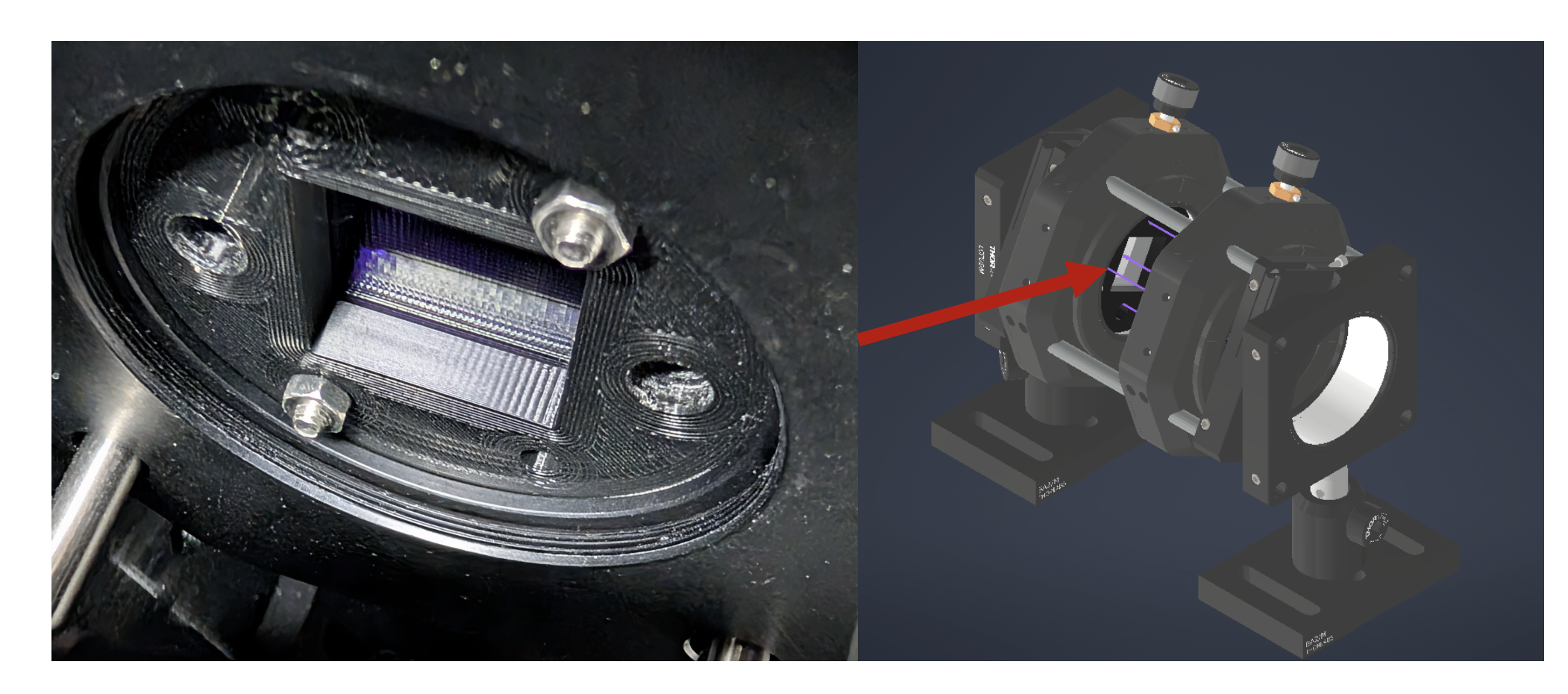}
    \caption{Left: 3D printed microlens array holder with holes for Thorlabs cage rods to fix angular alignment. Right: CAD model of the lenslet array IFU.}
    \label{fig:lla}
\end{figure}

The lenslet array IFU follows the BIGRE design \cite{Claudi2011OpticalSPHERE}. In this format, an image of the source is formed on the front surface of one lenslet array, which both splits the field and re-images the pupil into an array of micro-pupils. The second lenslet array demagnifies this image, while also imaging the micro-pupils formed by the first lenslet array at infinity to produce a telecentric output. An image of the lenslet array IFU is shown in figure~\ref{fig:lla}. The lenslet arrays are high-quality off-the-shelf components manufactured by SUSS micro-optics. They are seated in 3D-printed mounts and mounted on X-Y stages which are used to optically co-align the lenslets. The lenslet arrays are inclined at an angle of 26.6$^\circ$, which gives the largest possible vertical spacing between spectra. The lenslet array optical parameters are given in table~\ref{tab:lenslet_array_properties}. To prevent overlap between adjacent spectra on the detector, a bandpass filter with a central wavelength of $800$ nm and a FWHM of 65 nm was used.

\begin{table}[h]
\caption{Lenslet array IFU properties.}
\label{tab:lenslet_array_properties}
\centering
\begin{tabular}{ll}
\hline
\textbf{Parameter} & \textbf{Value} \\
\hline
Spaxel format & Up to $40 \times 40$ spaxels \\
Spaxel size & $0.5~\mathrm{mm} \times 0.5~\mathrm{mm}$  \\
Lenslet geometry & Plano-convex, square \\
Exit pupil position & $\infty$, telecentric \\
Exit slit sampling & 3 pix \\
R & $\approx$ 100 \\
Spectral Range & 777 - 835 nm \\
\hline
\end{tabular}
\end{table}

\section{Experiment}

\begin{figure}
    \centering
    \includegraphics[width=0.9\linewidth,trim= 0cm 0cm 4.2cm 0cm]{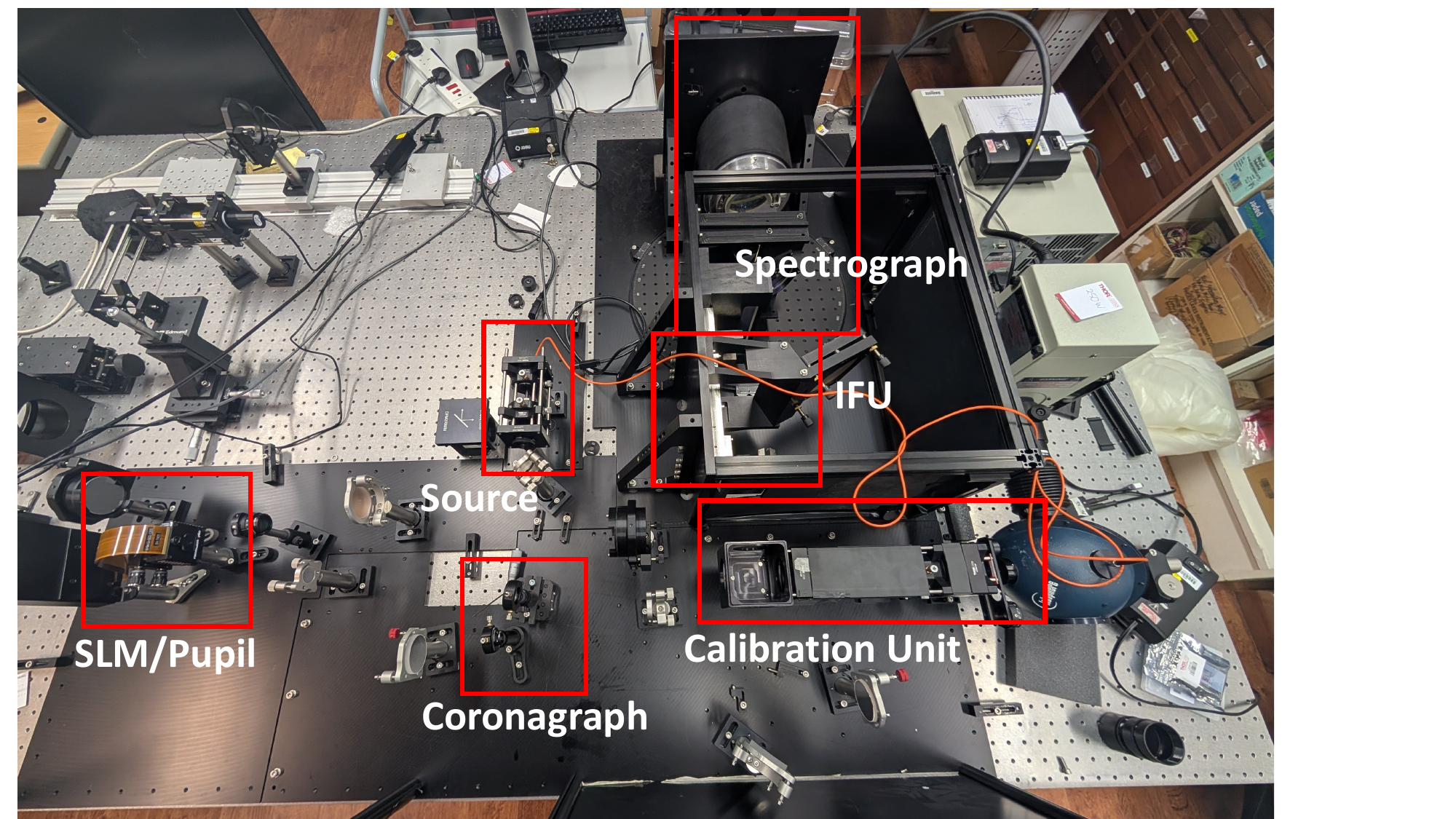}
    \caption{Labelled image of the \gls{ifs} test bench.}
    \label{fig:experiment}
\end{figure}

An image of the full experiment aligned on the optical bench is shown in figure~\ref{fig:experiment} with each subsystem highlighted in red. The principle behind the test bench is to simulate a high-contrast imaging system and provide the most realistic conditions under which the \glspl{ifu} can be compared. This section will provide a more detailed discussion of the design and operation of each subsystem.

\subsection{Pre-optics}

\begin{figure}
    \centering
    \includegraphics[width=\linewidth]{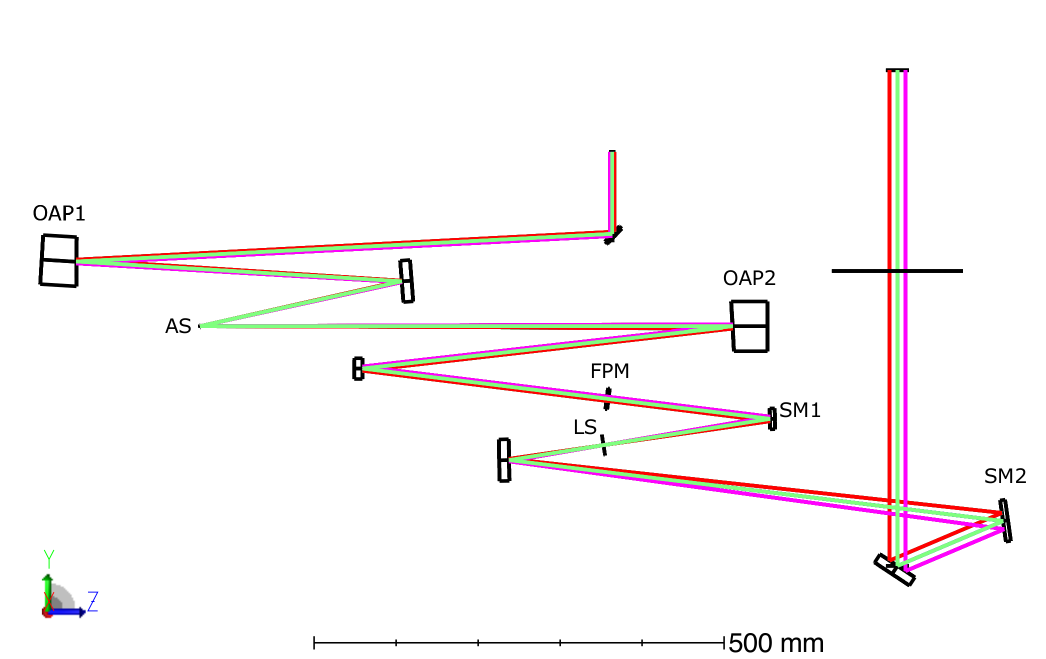}
    \caption{Achromatic pre-optics optical design. The main features in this new design are: a 1:1 imaging relay with two $\lambda/8$ surface quality off-axis parabolic mirrors (OAPs). AS, FPM and LS are the SLM surface, focal plane occulting mask and Lyot stop respectively. The OAP2/SM1 system acts as a beam reducer, and SM2 re-images the pupil at infinity to generate a telecentric output. Unlabelled elements are fold mirrors.}
    \label{fig:preoptics}
\end{figure}

The recently updated pre-optics aims to image a point source onto the integral field unit, with the \gls{slm} generating residual adaptive optics phase screens at the pupil, and a Lyot coronagraph set up in intermediate focal and pupil planes. The optical layout of the pre-optics system is shown in figure~\ref{fig:preoptics}. The pre-optics are fed either by a fibre which collects light from an integrating sphere, or a diverging beam from a HeNe alignment laser. The beam is collimated and reflects off of the \gls{slm} surface, which is then stopped down to meet the output focal ratio requirement. The Lyot coronagraph follows the \gls{slm}. The Lyot stop lies in the front focal plane of the final concave mirror to produce a telecentric output which is imaged onto the front surface of the IFU. The IFU output is then collimated and passed through a grating or prism. A camera barrel and 2k CCD detector then image the full dispersed \gls{ifu} field. The source is a 50 $\mu$m multimode optical fibre fed by an integrating sphere and 250W QTH lamp. The optics produce a f/842, diffraction-limited, telecentric beam conjugate to the \gls{ifu} surface, close to the f/850 nominal specification for the slicer. The overall system magnification is 3.75$\times$. The pupil stop on the \gls{slm} matches the size of the re-imaged pupil of the \gls{ghost} test bench, to allow future interoperability.  The \gls{slm} requires linearly polarised light to operate in a phase-only mode. We therefore added a polariser in the beam path prior to the \gls{slm} surface. To more accurately reflect a real telescope system and astronomical objects, we can optionally depolarise the light prior to transfer through the spectrograph using a rapidly rotating half-wave plate.

\subsection{Adaptive optics simulator}

Simulations of atmospheric residuals were implemented using a Meadowlark S19x12-500-1200 \gls{slm}. It was delivered factory-calibrated at 785 nm, and in this configuration the surface form is guaranteed to $\lambda/10$. An SLM was chosen to simulate wavefront deformations over a deformable mirror due to its smaller pixel size and lower cost. The pixel pitch of the \gls{slm} is 8 $\mu$m, which leads to approximately 420 pixels sampling the 3.35 mm pupil in the experiment. The \gls{slm} is stopped down to achieve the right focal ratio through use of a small 3D printed mask that fits over the display. The atmospheric turbulence simulations displayed on the \gls{slm} are implemented using analytical power spectral densities (PSDs) calculated for various \gls{ao} system configurations with the Astro-Tiptop package \cite{Neichel2021TIPTOP:PSF}. The \gls{ao} simulations in this paper made use of the SAXO example file, the single-conjugate \gls{ao} module which feeds SPHERE. The residual RMS wavefront error for this simulation was 83 nm. Given the PSD, a straightforward Monte Carlo method could be implemented to produce uncorrelated phase screens with the correct spatial scaling\cite{Schmidt2010}. Even with uncorrelated phase screens, we expect to see the effects of static aberrations \cite{Soummer2007SpeckleImages} interacting with the atmospheric speckles. However, quasi-static effects, which depend on the ratio of speckle lifetimes, are expected to be limited.

\subsection{Coronagraph}

The Lyot coronagraph focal plane mask is a 450 $\mu$m circular OD5 chrome dot on a 2~mm-thick fused silica glass plate. In the intermediate focal plane of the pre-optics, this corresponds to an angular diameter of $2.5 \lambda/D$ at the central wavelength 800~nm. An iris functions as a Lyot stop and is placed in the re-imaged pupil plane. The raw contrast of the coronagraph was analysed in HCIPy \cite{Por2018HighSimulator} according to the definition \cite{Douglas2018ReviewMetrics},

\begin{equation}
    C(\rho) = \frac{\eta_s (\rho)}{\eta_p(\rho)},
\end{equation}

where $\eta_p$ is the fraction of light contained in a $2 \lambda/D$ circular region centred on a position $\rho$ with the PSF at the same position, and $\eta_s$ is the encircled energy in that same region with the PSF centred on the coronagraph.

\begin{figure}[h]
    \centering
    \includegraphics[width=0.67\linewidth]{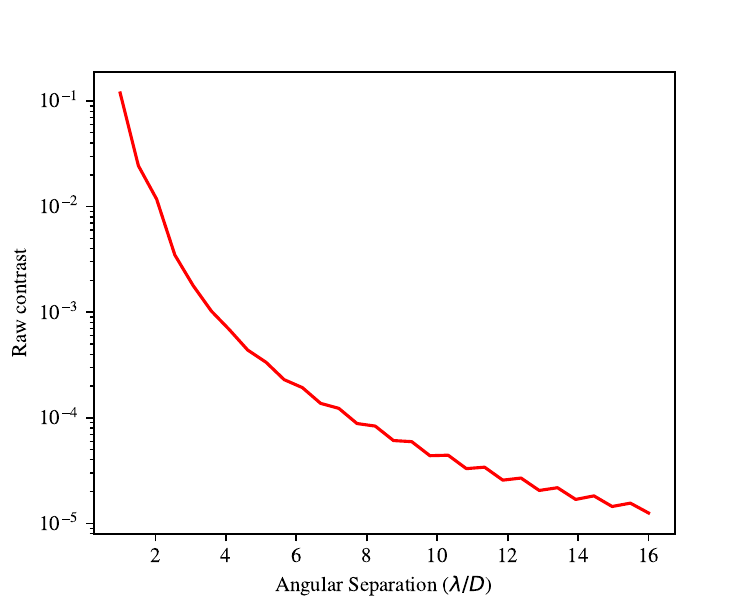}
    \caption{Raw contrast of the coronagraph as a function of radial distance from its centre. HCIpy simulation parameters such as stop sizes and lens focal lengths match the experiment. The simulation uses the Fraunhofer diffraction equation for focal to pupil plane transformations.}
    \label{fig:placeholder}
\end{figure}

\subsection{Other upgrades}
Compared to the previously reported design of the test bench\cite{Meyer2024AnIFUs}, we have upgraded the optomechanical model of the test bench to provide multiple improvements and advantages, in addition to those already mentioned above. Most changes were made with the objective of making the test bench a travelling instrument that could be taken to the GHOST facility at ESO, while also expanding on the modes of the test bench. These include:
\begin{enumerate}
    \item The disperser system is now mounted on a 2-axis rotary stage that allows for:
    \begin{itemize}
        \item Rotation of the disperser to any required angle to control the wavelength range, as well as flexibility to work with various kinds of dispersers that require different blaze angles, especially for those that are aimed at high-spectral resolution modes.
        \item The other axis allows clocking of the disperser with respect to the camera and CCD to align them and allow more efficient detector filling.
    \end{itemize}
    \item The camera optics and the CCD detector are now mounted on a rotary platform that allows for any camera angle with respect to the disperser.
    \item The overall optomechanical system was designed to be robust such that it allows for reliable and repeatable switching between the slicer and the lenslet \glspl{ifu}, with only one insertable component when changing the modes. The rest of the system, starting from the light source, pre-optics and calibration system, collimator and camera, is fixed.
    \item The pre-optics system has been designed with seamless integration in mind, and for use with either an SLM in the lab or the beam from GHOST.
    \item Accurate mounts for the OAPs and other optics were designed that allowed us to align all the optics to the required tolerances as well as maintain it when we transport the instrument to ESO.
\end{enumerate}
\subsubsection{Polarimetry with Slicer and Lenslet IFU}
One of the key technical requirements coming out of the Science Work Package of the ongoing \gls{pcs} R\&D Roadmap study is the need for a polarimetry mode in conjugation with \gls{ifu} spectroscopy. Integral field spectropolarimetry (IFSP) is a relatively untested technology. To explore its feasibility for PCS, as an extension of the ongoing tests, we have designed an optical and optomechanical system that can be integrated with the test bench to characterise the polarimetric behaviour of the system and test the limits for employing IFSP. Our designed system consists of multiple half-wave plates and high-efficiency polarisers in the pre-optics suitably placed to allow for creation of various kinds of polarised states of light in the system. From this we will create the instrument’s Mueller matrix to quantify the polarimetric behaviour of the system and come up with suitable calibration methods. While this mode of the instrument will be explored in more detail in future, the capability has already been implemented in the current version of the test bench.

\section{Calibration and Data Reduction}

Calibration data were collected using the calibration unit shown in figure~\ref{fig:experiment}. The calibration unit includes a large right-turning prism mirror that can be removed to allow the ``science" beam to pass into the \gls{ifs}. It is used to image a uniform flat field from the port of an integrating sphere, or a pinhole/calibration mask which is placed in front of the sphere port. Calibrations common to both \gls{ifu} configurations include:

\begin{itemize}
    \item Bias, dark corrections.
    \item A bad pixel mask generated using the IRAF CCD mask routine.
    \item Detector flat field using the 2-inch port on the integrating sphere with a QTH lamp source.
    \item Background subtraction to remove lab contamination.
    \item An IFU flat field using the QTH lamp source.
\end{itemize}

In addition, spectral and spatial calibrations were carried out for each \gls{ifu} due to their different architectures.

\subsection{Image Slicer}

\begin{figure}
     \centering
     \includegraphics[width=1\linewidth]{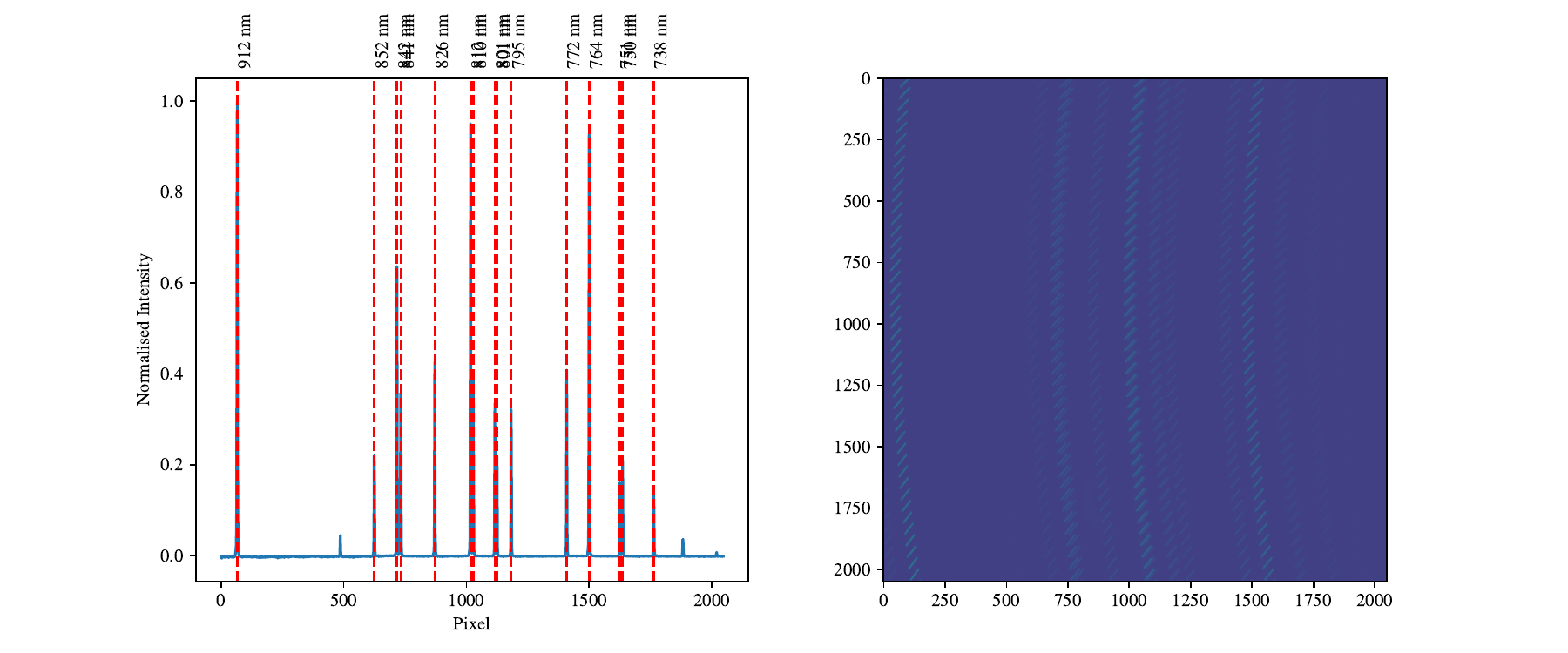}
     \caption{Argon spectrum from one row of the image slicer arc lamp calibration frame, after spectral fitting using the RANSAC algorithm.}
     \label{fig:spectral}
 \end{figure}

 The image slicer \gls{ifu} requires a spectral calibration and a geometric calibration. Wavelength calibrations were carried out by imaging the integrating sphere illuminated by an argon arc lamp. The argon lines produced satisfactory calibration frames for both the image slicer and lenslet array \glspl{ifu}. For the image slicer design, which is able to resolve the doublet lines, a RANSAC algorithm was used to automate spectral matching. An example of the identification of strong lines is given in figure~\ref{fig:spectral}. The spectral matching algorithm was inspired by the RASCAL algorithm \cite{Veitch-Michaelis2019RASCAL:Calibration}. To summarise, RANSAC is a computer vision algorithm for fitting with robust outlier rejection. To apply this to the problem of matching spectral lines, we sample two points without replacement from the population of unknown peaks detected in the spectrum. The argon lines and the linear dispersion of the spectrograph can then be used to fit the remaining peaks. An inlier threshold is set, and if it is met the best fit is updated. The algorithm runs a set number of times, or terminates early if it matches every peak. After peaks have been mapped to their transition lines, Gaussian curves can be fitted to the peaks to find their centres with sub-pixel accuracy. A simultaneous fitting is preferred to take into account cross-talk effects.

\begin{figure}
     \centering
     \includegraphics[width=0.8\linewidth]{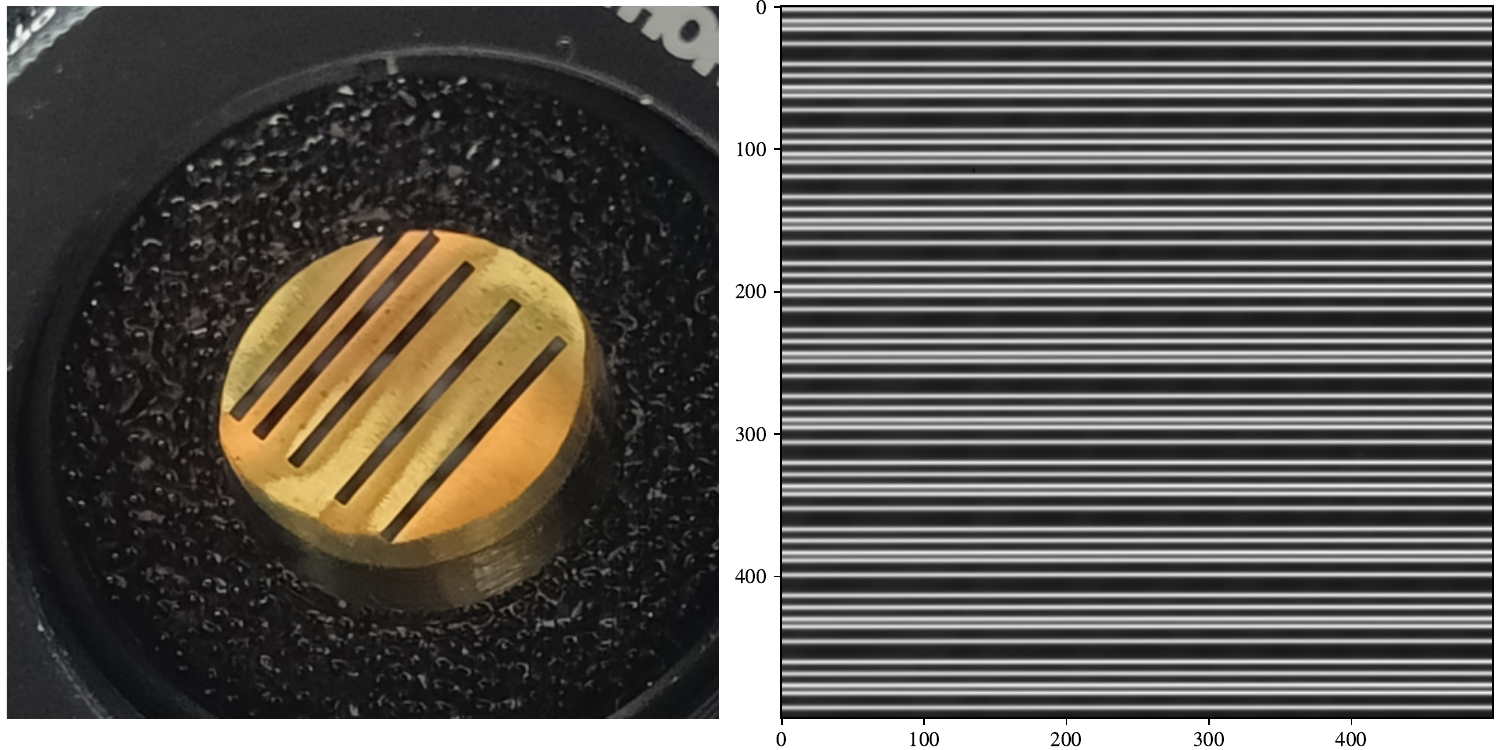}
     \caption{Image of an etched line mask (left) in its holder and an example of a dispersed image of the line mask after propagation through the slicer \gls{ifu}.}
     \label{fig:linemask}
 \end{figure}

The geometric calibration for the twisted image slicer uses a ``line mask" with precisely known line positions imaged from the calibration unit object plane onto the \gls{ifu}. The line masks were produced using in-house photolithography equipment at Oxford, and are etched on 25~$\mu$m thick brass sheets. The width of the lines is 100~$\mu$m, which, when imaged by the 2x magnification calibration unit, produces an image less than $^1/_3$ of the PSF FWHM at 800~nm on the slicer. By aligning the mask in the across-slice direction, a series of lines are produced across the detector. Each mirror will have $N$ lines at known spatial separations, allowing an $N$-order polynomial fit to interpolate pixel positions on the slicing mirror. In this way, it is possible to assign a 2D spatial coordinate on the image slicer (a spaxel) to each pixel on the detector. An image of an example line mask, and a sub-region of a resulting dispersed line mask image on the CCD are given in figure~\ref{fig:linemask}.

 \begin{figure}
     \centering
     \includegraphics[width=0.95\linewidth]{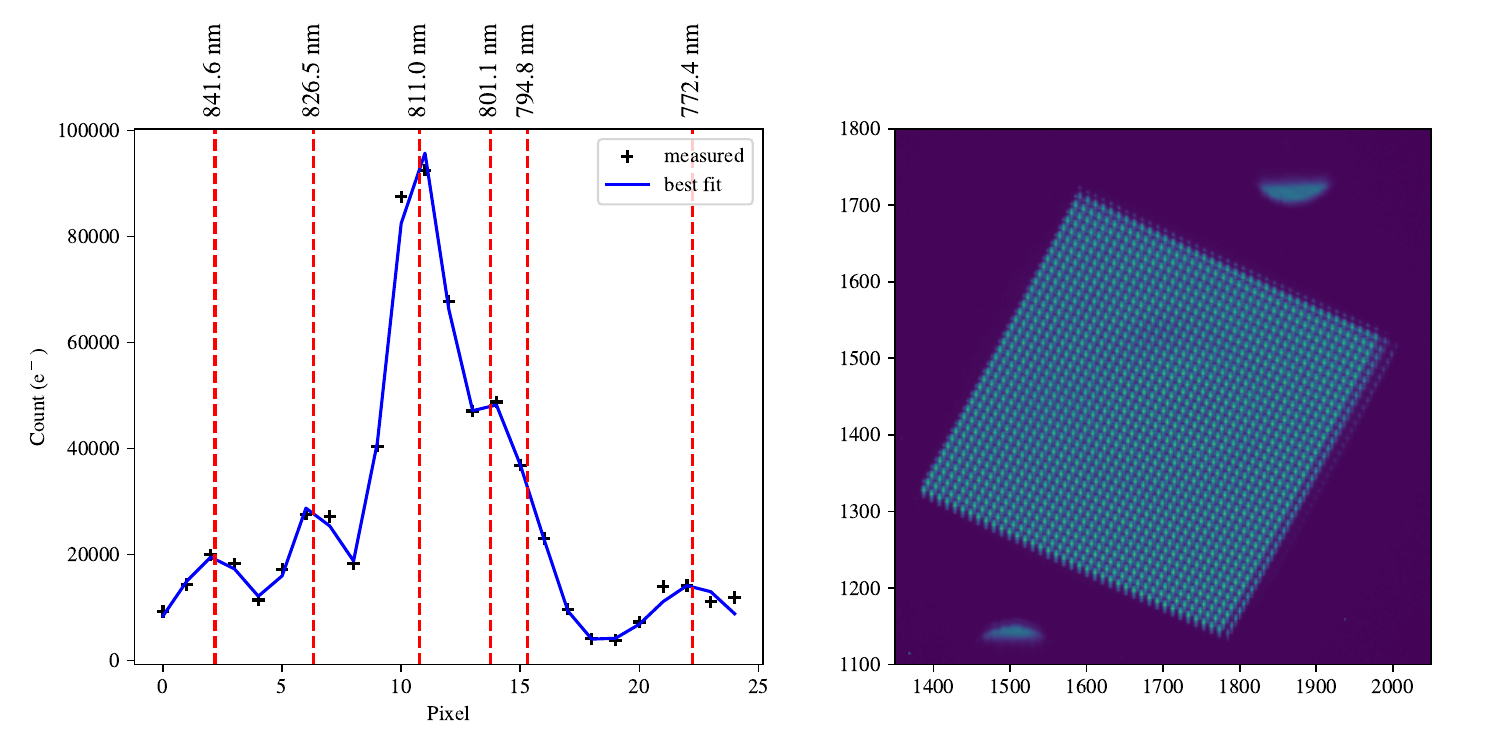}
     \caption{Argon spectrum and best fit calculated via least squares for a single lenslet spaxel (left) and an image of the corresponding argon flat field image (right).}
     \label{fig:spectral_lenslet}
 \end{figure}

\subsection{Lenslet Array}

The lenslet array geometric calibration is made by generating a model continuum spectrum in Zemax and performing a correlation to find the location of each lenslet image. Each pixel inside a box of $5\times 20$ pixels centred on this point is assigned an $(x, y)$ spaxel coordinate. The spectral calibration is, however, less straightforward than for the image slicer. This is due to the significant overlap between argon spectral lines as a result of the low spectral resolving power of the lenslet \gls{ifs}. The current method closely follows that used for GPI \cite{Macintosh2014FirstImager} in which 2D Gaussian functions are fit simultaneously to each spectral line in a given spaxel.

\subsection{Data cube interpolation}

\begin{figure}[h]
    \centering
    \includegraphics[width=0.95\linewidth]{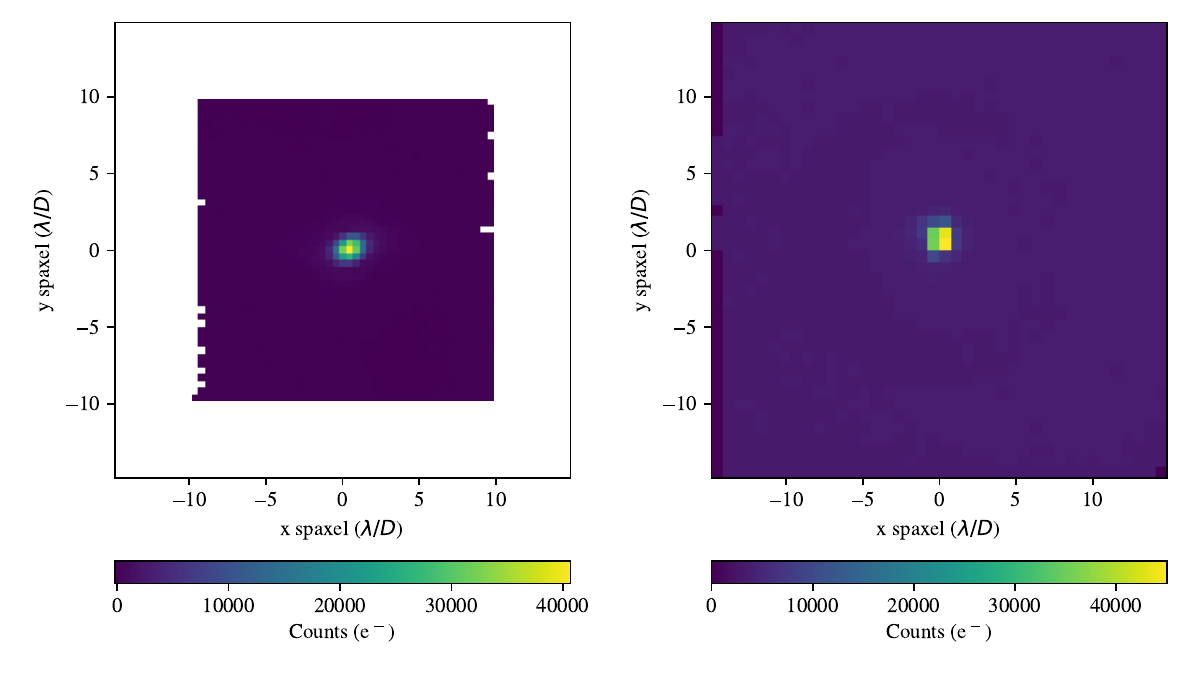}
    \caption{Data cubes generated from a point source imaged through the slicer (left) and lenslet array (right) \glspl{ifu}. The data cubes have been collapsed to a 2D image through averaging over the spectral dimension. The units of the spaxels are given in $\lambda/D$ units at 800~nm. The extent of the slicer image has been modified to match the field size of the lenslet array to indicate the wider overall field of view.}
    \label{fig:cubes}
\end{figure}

To construct the data cubes used to evaluate contrast in this paper, four inputs were required: a reduced, calibrated image of a continuum point source, the associated uncertainty in the photoelectron count and the geometric and spectral pixel maps. Using these data, each pixel value is associated with a wavelength and a location on the \gls{ifu} focal plane such that the intensity can be written $f(x, y, \lambda)$, where $\lambda$ is the wavelength, $x$ denotes the along-slice coordinate, or simply one of the lenslet axes, and $y$ the across-slice coordinate, or the orthogonal lenslet axis. Similarly, the uncertainties can be written $\sigma(x, y, \lambda)$. This function is then interpolated onto a linear grid of integer spaxel positions. In the lenslet array case, linear interpolation is necessary only over the spectral direction, and the uncertainties are propagated through the same formula for each lenslet. For the image slicer,  a combination of Delaunay triangulation and barycentric linear interpolation is used to interpolate the function $f$ over the continuous variables $x$ and $\lambda$. The $y$-axis values are already sampled on a regular grid. Using barycentric linear interpolation also enables propagation of uncertainty into the interpolated, regular spaxel grid. The uncertainties for the three bounding pixels computed by the triangulation are simply weighted by the barycentric weights and added in quadrature.

\begin{equation}
    \sigma_{f, \mathrm{spax}} = \sqrt{\sum_{i=1}^3{\left( b_{i} \sigma(x_i, y, \lambda_i)\right)}^2 },
\end{equation}

where the $b_i$ are the barycentric weights of the function points $f(x_i, y, \lambda_i)$ used in the interpolation, and $\sigma(x_i, y, \lambda_i)$ are the associated uncertainties. If we consider only the uncertainty due to detector noise (i.e. background, dark and readout noise), and we assume that the noise in each pixel is identical and independent, we can propagate the detector noise through the interpolation pipeline to find the noise floor in the measurement. Spectrally collapsed data cubes are shown in figure~\ref{fig:cubes} and are scaled to indicate the different field sizes. The spatial dimension is undersampled on the lenslet array \gls{ifu}, with the FWHM of the PSF sampled by  $1.3$ pixels at the central wavelength. This is due to the 500~$\mu$m lenslet pitch, which enables a wider spectral bandpass and higher resolving power than the previous design reported in Ref.~\citenum{Meyer2024AnIFUs} which used the $300~\mu$m lenslets. It should also be noted that the slicer image is averaged over a wider wavelength range, which will cause the PSF to appear wider.

\section{Initial Contrast Results}

So far, only the slicer \gls{ifu} has been characterised using the full pre-optics system, including the \gls{ao}  residuals and the coronagraph. The lenslet array characterisation was made using only the calibration unit with a pinhole in the object plane. The lenslet \gls{ifu} data was also taken with a different CCD camera and background noise conditions. A fair comparison  will therefore require controlling for the differential detector noise. To do so, the dark, background and readout noise were measured from calibration data, and the associated uncertainties propagated through the reconstruction process. Some simple assumptions are made, including that the detector and background noise is the same in each pixel, and that noise in each pixel is independent. The median propagated detector noise in the data cube is then taken as the noise floor, scaled by a factor of five for plotting on the graph. To measure the contrast, we use the method adopted for SPHERE \cite{Mesa2015PerformanceSPHERE}. In summary, a box of size $1.5 \lambda /D$ is used to measure the standard deviation of pixel intensities. This is then radially averaged and normalised by the peak of the PSF taken from a Gaussian fit. For the coronagraph case, the peak intensity is taken from the unmodified PSF.

\begin{figure}[h]
    \centering
    \includegraphics[width=\linewidth]{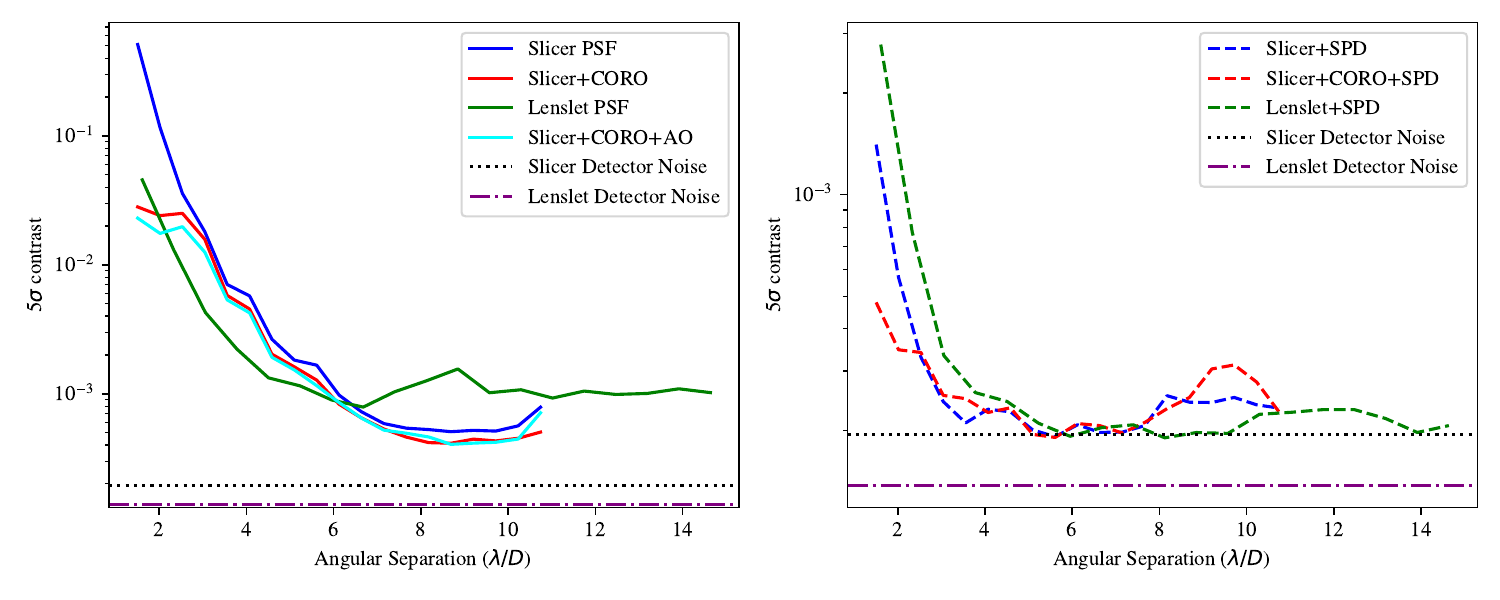}
    \caption{Initial contrast curves achieved with both configurations. The left-hand plot compares the image slicer measurements with the lenslet array measurements without post-processing. The slicer results include the coronagraph and adaptive optics residuals which are labelled CORO, AO in the legend. The lenslet array \gls{ifu} results include only the unmodified PSF. The detector noise floor is indicated by the horizontal dotted lines. On the right-hand side, the slicer and lenslet array results are compared after post-processing with spectral deconvolution.}
    \label{fig:contrast}
\end{figure}

From figure~\ref{fig:contrast}, up to $6 \lambda/D$ the lenslet array \gls{ifu} provides a deeper contrast than the slicer in the ``PSF only'' case. Beyond this, the slicer \gls{ifu} achieves a deeper contrast over the remaining shared field-of-view. The effects of the coronagraph and adaptive optics were limited, although these measurements were made with the Lyot stop open to allow enough photons through and so do not represent the nominal operation of the coronagraph. Furthermore, speckles generated by the adaptive optics were visually buried in the detector noise, such that only one or two diffraction rings were visible. To test the effects of spectral post-processing techniques, we applied spectral deconvolution \cite{Thatte2007VeryDeconvolution} to the data. In short, this method involves scaling the image such that the PSF does not change size with increasing wavelength, which causes an off-axis companion to move through the image. The images are then recombined in such a way that the PSF is subtracted, but the planet is not as it represents a high-frequency modulation on top of slow variations in pixel intensity caused by the PSF wavelength scaling. The slicer benefits more from spectral deconvolution, as the contrast touches the detector noise floor. The lenslet \gls{ifu}, however, remains at least 2$\times$ higher than its noise floor value. The magnitude of the contrast gain is also greater, particularly at small angular separations. For example, at 4 $\lambda/D$ the slicer gains 24$\times$ compared to only $7\times$ for the lenslet array. The improved performance with spectral deconvolution is likely due to the increased spectral resolving power enabling a better PSF fit.

\section{Conclusion and further work}

This paper presents an upgraded design of the \gls{ifs} test bench at Oxford, in addition to its current suite of calibrations and some initial results comparing the contrast of the prototype twisted image slicer and a new lenslet array \gls{ifu}. The test bench now simulates a high-contrast imaging system through a Lyot coronagraph and adaptive optics residuals injected by an \gls{slm}. Using this equipment, a point source is imaged onto each \gls{ifu} to produce a diffraction-limited PSF around which the $5\sigma$ contrast level can be measured. For the image slicer and lenslet array \glspl{ifu}, spectral calibrations were made using an argon arc lamp. Spectral lines were fully resolved for the image slicer, and an automated RANSAC algorithm was used to match them to NIST data row-by-row. For the lenslet array, 2D Gaussians were fit to six visible spectral lines simultaneously using least squares enabled by a dual-annealing optimiser. The spatial calibration for the image slicer \gls{ifu} was made using a precision line mask fabricated in-house. The six lines on the mask enabled a sixth-order polynomial to be used to interpolate the spaxel positions of each pixel along the slices. With the calibrations made, the $5\sigma$ contrast of two \glspl{ifu} were compared. For now, the lenslet array measurements are limited to the unmodified PSF only as the beam relay system was not yet ready. The dark, background and readout noises were propagated through the data cube reconstruction for each of the \glspl{ifu}. The resulting detector noise floor was then plotted for each contrast curve, and it represents a fundamental limit on the contrast that could be achieved for each measurement. The results show that the lenslet array \gls{ifu} outperforms the image slicer in contrast up to about $6\lambda/D$, after which the slicer provides a deeper contrast. It was also found that spectral deconvolution was more effective on the image slicer, likely due to its higher spectral resolving power. It experienced approximately a 3.4$\times$ greater gain at 4$\lambda/D$, and was able to bring the contrast down to the noise floor at $5\lambda/D$. The slicer outperformed the lenslet array at small angular separations after post-processing.

Based on these initial results, there are a number of improvements and further tests that could be implemented on the test bench in the next few months:

\begin{itemize}
    \item The source brightness might be increased using a supercontinuum laser to ensure that we are not detector-noise dominated.
    \item An improved wavelength calibration will be made for the lenslet array \gls{ifu} with a monochromator.
    \item A full test will be carried out with the coronagraph and adaptive optics residuals for the lenslet array \gls{ifu}.
    \item Measurements of the long-term stability of the contrast curve will be made to identify the influence of quasi-static contributions.
    \item Adaptive optics residuals will be tested across a number of atmospheric cases, in addition to using correlated phase screens calculated from Monte Carlo simulations.
    \item The \gls{ifs} will travel to ESO Headquarters in Garching for integration and repetition of these tests behind the GHOST cascade adaptive optics simulator.
\end{itemize}

\acknowledgments 

The authors would like to acknowledge the mechanical workshop for producing the experiment infrastructure, Paul Pattinson for etching of lithographic masks and Rick Makin for coatings. The authors would also like to acknowledge the UKRI grant funding for this project (ST/Y005392/1).

\bibliography{references} 

\begin{thebibliography}{10}

\bibitem{Kasper2021PCSELT}
Kasper, M. and {others}, ``{PCS — A Roadmap for Exoearth Imaging with the ELT},'' {\em The Messenger}~{\bf 182},  38--43 (3 2021).

\bibitem{Beuzit2019SPHERE:Telescope}
Beuzit, J.-L. et~al., ``{SPHERE: the exoplanet imager for the Very Large Telescope},'' {\em Astronomy {\&} Astrophysics}~{\bf 631},  A155 (11 2019).

\bibitem{Macintosh2014FirstImager}
Macintosh, B. et~al., ``{First light of the Gemini Planet Imager},'' {\em Proceedings of the National Academy of Sciences}~{\bf 111},  12661--12666 (9 2014).

\bibitem{Currie2020On-skySystem}
Currie, T.~M. et~al., ``{On-sky performance and recent results from the Subaru coronagraphic extreme adaptive optics system},'' in [{\em Adaptive Optics Systems VII}{\nolinebreak\hspace{0.1em}]},  Schmidt, D., Schreiber, L., and Vernet, E., eds.,  330, SPIE (12 2020).

\bibitem{Absil2024METISTesting}
Absil, O. et~al., ``{METIS high-contrast imaging: from final design to manufacturing and testing},'' (2024).

\bibitem{Houlle2021DirectModule}
Houll{\'{e}}, M., Vigan, A., Carlotti, A., Choquet, E., Cantalloube, F., Phillips, M.~W., Sauvage, J.-F., Schwartz, N., Otten, G. P. P.~L., Baraffe, I., Emsenhuber, A., and Mordasini, C., ``{Direct imaging and spectroscopy of exoplanets with the ELT/HARMONI high-contrast module},'' {\em Astronomy {\&} Astrophysics}~{\bf 652},  A67 (8 2021).

\bibitem{Haffert2020Multi-coreFirst-light}
Haffert, S.~Y. et~al., ``{Multi-core fibre-fed integral-field unit (MCIFU): overview and first-light},'' (2020).

\bibitem{Marconi2024ANDESDevelopments}
Marconi, A. et~al., ``{ANDES, the high resolution spectrograph for the ELT: science goals, project overview, and future developments},'' (2024).

\bibitem{Stelter2023TheSpectrograph}
Stelter, R.~D., Kupke, R., Skemer, A.~J., Sallum, S., Bourgenot, C., and Martinez, R.~A., ``{The SCALES slenslit: a unique exoplanet spectrograph},'' (2023).

\bibitem{Meyer2024AnIFUs}
Meyer, R.~E., Tecza, M., Thatte, N.~A., and Sukegawa, T., ``{An integral field unit for the planetary camera and spectrograph (ELT-PCS): comparing lenslet and image slicer-based IFUs},'' in [{\em Advances in Optical and Mechanical Technologies for Telescopes and Instrumentation VI}{\nolinebreak\hspace{0.1em}]},  Navarro, R. and Jedamzik, R., eds.,  61, SPIE (8 2024).

\bibitem{Tecza2006SWIFT:Spectrograph}
Tecza, M., Thatte, N., Clarke, F., Goodsall, T., and Symeonidis, M., ``{SWIFT: An adaptive optics assisted I/z band integral field spectrograph},'' (2006).

\bibitem{Engler2022GPU-basedTestbench}
Engler, B., Kasper, M., Bristow, P., Heritier, C.~T., Leveratto, S., Verinaud, C., Le~Louarn, M., Nousiainen, J., Helin, T., Bonse, M., Quanz, S., Glauser, A., Bernard, J., Gratadour, D., and Clare, R., ``{GPU-based High-order adaptive OpticS Testbench},'' (2022).

\bibitem{Tecza2022ImageELT-PCS}
Tecza, M., Meyer, E., Sukegawa, T., Nakayasu, T., and Koyama, M., ``{Image slicing with a twist: design and manufacturing of a prototype image slicer for ELT-PCS},'' in [{\em Advances in Optical and Mechanical Technologies for Telescopes and Instrumentation V}{\nolinebreak\hspace{0.1em}]},  Geyl, R. and Navarro, R., eds.,  98, SPIE (8 2022).

\bibitem{Tecza2014ImageOptics}
Tecza, M., ``{Image slicing with a twist: spatial and spectral Nyquist sampling without anamorphic optics},'' in [{\em Advances in Optical and Mechanical Technologies for Telescopes and Instrumentation}{\nolinebreak\hspace{0.1em}]},   {\bf 9151} (2014).

\bibitem{Claudi2011OpticalSPHERE}
Claudi, R. et~al., ``{Optical design and test of the BIGRE-based IFS of SPHERE},''  81671S (9 2011).

\bibitem{Neichel2021TIPTOP:PSF}
Neichel, B., Beltramo-Martin, O., Plantet, C., Rossi, F., Agapito, G., Fusco, T., Carolo, E., Carla, G., Cirasuolo, M., and Van Der~Burg, R., ``{TIPTOP: a new tool to efficiently predict your favorite AO PSF},'' (2021).

\bibitem{Schmidt2010}
Schmidt, J.~D.,  [{\em {Numerical simulation of optical wave propagation: With examples in MATLAB}}{\nolinebreak\hspace{0.1em}]}, SPIE (2010).

\bibitem{Soummer2007SpeckleImages}
Soummer, R., Ferrari, A., Aime, C., and Jolissaint, L., ``{Speckle Noise and Dynamic Range in Coronagraphic Images},'' {\em The Astrophysical Journal}~{\bf 669},  642--656 (11 2007).

\bibitem{Por2018HighSimulator}
Por, E.~H., Haffert, S.~Y., Radhakrishnan, V.~M., Doelman, D.~S., van Kooten, M., and Bos, S., ``{High Contrast Imaging for Python (HCIPy): an open-source adaptive optics and coronagraph simulator},'' (2018).

\bibitem{Douglas2018ReviewMetrics}
Douglas, E., Zimmerman, N., Ruane, G., Mazoyer, J., Riggs, A.~E., Carlomagno, B., Huby, E., Fogarty, K., Por, E., Absil, O., Baudoz, P., Galicher, R., Beaulieu, M., Cady, E.~J., Carlotti, A., Doelman, D., Guyon, O., Haffert, S., Jewell, J.~B., Jovanovic, N., Keller, C., Kenworthy, M.~A., Knight, J., Kuhn, J., Miller, K., N’Diaye, M., Pueyo, L., Sirbu, D., Snik, F., Wallace, J.~K., Wilby, M., and Ygouf, M., ``{Review of high-contrast imaging systems for current and future ground- and space-based telescopes I: coronagraph design methods and optical performance metrics},'' in [{\em Space Telescopes and Instrumentation 2018: Optical, Infrared, and Millimeter Wave}{\nolinebreak\hspace{0.1em}]},  MacEwen, H.~A., Lystrup, M., Fazio, G.~G., Batalha, N., Tong, E.~C., and Siegler, N., eds.,  98, SPIE (8 2018).

\bibitem{Veitch-Michaelis2019RASCAL:Calibration}
Veitch-Michaelis, J. and Lam, M.~C., ``{RASCAL: Towards automated spectral wavelength calibration},'' (12 2019).

\bibitem{Mesa2015PerformanceSPHERE}
Mesa, D., Gratton, R., Zurlo, A., Vigan, A., Claudi, R.~U., Alberi, M., Antichi, J., Baruffolo, A., Beuzit, J.-L., Boccaletti, A., Bonnefoy, M., Costille, A., Desidera, S., Dohlen, K., Fantinel, D., Feldt, M., Fusco, T., Giro, E., Henning, T., Kasper, M., Langlois, M., Maire, A.-L., Martinez, P., Moeller-Nilsson, O., Mouillet, D., Moutou, C., Pavlov, A., Puget, P., Salasnich, B., Sauvage, J.-F., Sissa, E., Turatto, M., Udry, S., Vakili, F., Waters, R., and Wildi, F., ``{Performance of the VLT Planet Finder SPHERE},'' {\em Astronomy {\&} Astrophysics}~{\bf 576},  A121 (4 2015).

\bibitem{Thatte2007VeryDeconvolution}
Thatte, N., Abuter, R., Tecza, M., Nielsen, E.~L., Clarke, F.~J., and Close, L.~M., ``{Very high contrast integral field spectroscopy of AB Doradus C: 9-mag contrast at 0.2 arcsec without a coronagraph using spectral deconvolution},'' {\em Monthly Notices of the Royal Astronomical Society}~{\bf 378}(4) (2007).

\end{thebibliography}
\bibliographystyle{spiebib}  

\end{document}